\documentstyle[aps,pra,twocolumn]{revtex}
\begin{document}
\draft

\newcommand{\pp}[1]{\phantom{#1}}
\newcommand{\be}{\begin{eqnarray}}
\newcommand{\ee}{\end{eqnarray}}
\newcommand{\ve}{\varepsilon}
\newcommand{\vs}{\varsigma}
\newcommand{\Tr}{{\,\rm Tr\,}}
\newcommand{\pol}{\frac{1}{2}}

\title{
Nonlinear von Neumann-type equations: Darboux invariance and spectra 
}
\author{Maciej Kuna$^1$, Marek~Czachor$^{1,2}$, and Sergiej~B.~Leble$^1$}
\address{
$^1$ Wydzia{\l} Fizyki Technicznej i Matematyki Stosowanej\\
Politechnika Gda\'{n}ska,
ul. Narutowicza 11/12, 80-952 Gda\'{n}sk, Poland\\
$^2$ Arnold Sommerfeld Institute for Mathematical Physics\\
Technical University of Clausthal, 38678 Clausthal-Zellerfeld,
Germany
}
\maketitle

\begin{abstract}
Generalized Euler-Arnold-von Neumann density matrix equations 
can be solved by a binary Darboux transformation, given here in a new form: 
$\rho[1]=e^{P\ln(\mu/\nu)}\rho e^{-P\ln(\mu/\nu)}$, where
$P=P^2$ is explicitly constructed in terms of conjugated Lax pairs, 
and $\mu$, $\nu$ are complex. As a result spectra of 
$\rho$ and $\rho[1]$ are identical. 
Transformations allowing to shift and rescale spectrum of a solution
are introduced, and a class of stationary seed solutions is discussed.
\end{abstract}

\narrowtext
\section{Introduction}

In this Letter we want to address the following problem. Assume
we have a nonlinear von Neumann equation (vNE) whose solutions can be
generated by a binary Darboux transformation \cite{LU,ZL,L,U}.
The question is what is the relation between spectrum of a ``seed"
solution $\rho$ and this of the Darboux-transformed $\rho[1]$.
The issue is especially
important if $\rho[1]$ is to be a density matrix: Its spectrum
cannot be negative, but does the fact that we start with a
positive $\rho$ imply positivity of $\rho[1]$? 

Quite recently 
density matrix solutions of the nonlinear Euler-Arnold-von Neumann equation 
(EAvNE) $i\dot\rho=[H,\rho^2]$ 
were constructed by this technique in \cite{SLMC}. The solutions
$\rho[1]$ were obtained from the seed $\rho$ satisfying
$[\Delta_a,H]=0$, $a\in \bbox R$, with
nontrivial $\Delta_a:=\rho^2-a\rho$. The explicit form was
shown to be
\be
\rho[1](t)
&=&
e^{-iaHt}\Big(\rho(0)+(\mu-\bar\mu)F_a(t)^{-1}\nonumber\\
&\pp =&
\times e^{-\frac{i}{\mu}\Delta_a t}
\big[|\varphi(0)\rangle\langle\varphi(0)|,H\big]
e^{\frac{i}{\bar \mu}\Delta_a t}\Big)e^{iaHt},\label{rho[1]}
\ee
where $\rho(0)$ is an initial condition for $\rho$, $\mu$ is a
complex parameter of the Darboux transformation,
$|\varphi(0)\rangle$ a solution of the Lax pair at $t=0$, and
$$F_a(t)=
\langle\varphi(0)|
\exp\Big(
i
\frac{\mu-\bar\mu}{|\mu|^2}\Delta_a t
\Big)|\varphi(0)\rangle.$$ 
An analysis of explicit examples (e.g., for $H$ corresponding to 
a 1-dimensional harmonic
oscillator) revealed several interesting and unexpected  properties 
of $\rho[1](t)$. For example, in all the examples the dynamics
was found to be asymptotically {\it linear\/}, but with different asymptotics
for $t\to+\infty$ and $t\to-\infty$ and a kind of ``phase
transition" around $t=0$. Also in all the examples spectrum of
$\rho[1](t)$ was equal to this of $\rho(0)$. It is difficult to see these
properties in the formula (\ref{rho[1]})
and therefore it was not clear to what extent they followed from
the choice of the seed $\rho$, and whether they are a feature of only this
particular nonlinear equation. 

In this Letter we prove that spectra of $\rho[1]$ and $\rho$ coincide
whenever the 
equation they satisfy can be obtained as consistency condition
for a Lax pair which is covariant under the {\it binary\/}
Darboux transformation. The result is therefore not a pecularity of
the EAvNE and does not depend on the choice of seed solutions.
We shall give a general proof which does
not refer to any concrete equation but the 
notion of the binary-Darboux covariance will be illustrated on a
class on nonlinear vNE's. Finally we shall show how
to shift and rescale the spectra. Such properties are important in
order to generate normalized density matrices $\rho[1]$ from $\rho$'s
that are either nonpositive or nonnormalized.

\section{Zakharov-Shabat problem and Darboux-type transformations}

Let $V$ and $J$ be linear operators, $s$ a parameter, and $\mu$ a
complex number. We shall consider two types of Zakharov-Shabat (ZS)
problems. Let $|\varphi\rangle \in {\cal H}$ be a vector from a
Hilbert space.  The Hilbert-space ZS equation is
\be
i\partial_s|\varphi\rangle 
=
\big(V-\mu J)|\varphi\rangle. \label{ket}
\ee
The second type of equation is obtained by taking a
``second-quantized" problem
\be
i\partial_s \hat\varphi
=
\big(V-\mu J)\hat\varphi\label{hat}
\ee
where $\hat \varphi$ is an operator. 

In order to introduce the binary Darboux transformation of a ZS
problem we supplement (\ref{ket}) or (\ref{hat}) by an
appropriate {\it pair\/} of {\it conjugate\/} ZS equations:
\be
-i\partial_s\langle\chi|
&=&
\langle\chi|\big(V-\nu J), \label{bra1}\\
-i\partial_s\langle\psi|
&=&
\langle\psi|\big(V-\lambda J), \label{bra2}
\ee
for (\ref{ket}), and 
\be
-i\partial_s\hat\chi
&=&
\hat\chi\big(V-\nu J), \label{hat1}\\
-i\partial_s\hat\psi
&=&
\hat\psi\big(V-\lambda J) \label{hat2}
\ee
for (\ref{hat}). We define the operator $P$ by
\be 
P&=&\frac{|\varphi\rangle\langle\chi|}{\langle\chi|\varphi\rangle}\label{P1}
\\
P&=&\hat\varphi(p\hat\chi\hat\varphi p)^{-1}\hat\chi\label{P2}
\ee
for the two cases, respectively, where the operator $p$ satisfies $p^2=p$ and
$\partial_s p=0$. The inverse in (\ref{P2}) is understood as the
one in a $p$-invariant subspace: For any operator $\hat x=p\hat
x p$ its inverse $\hat x^{-1}=p\hat x^{-1}p$ satisfies $\hat
x\hat x^{-1}=\hat x^{-1}\hat 
x=p$. In both cases $P$ is an idempotent. For example,
\be
P^2 &=&\hat\varphi(p\hat\chi\hat\varphi p)^{-1}\hat\chi
\hat\varphi(p\hat\chi\hat\varphi p)^{-1}\hat\chi\nonumber\\
&=&
\hat\varphi(p\hat\chi\hat\varphi p)^{-1}p\hat\chi
\hat\varphi p(p\hat\chi\hat\varphi p)^{-1}\hat\chi=
\hat\varphi p(p\hat\chi\hat\varphi p)^{-1}\hat\chi=P.
\nonumber
\ee
Define
\be
V[1] = V + (\mu-\nu)[P,J]
\ee
and 
\be
\langle\psi[1]|=\langle\psi|\Big(\bbox 1
- \frac{\nu-\mu}{\lambda-\mu}P\Big)\label{BDT}\\
\hat\psi[1]=\hat\psi\Big(\bbox 1
- \frac{\nu-\mu}{\lambda-\mu}P\Big).
\ee
The following theorem can be proved by a direct calculation 
(for a simple proof see \cite{SLMC})

\medskip\noindent
{\bf Theorem~1.} 
\be
-i\partial_s\langle\psi[1]|
&=&
\langle\psi[1]|\big(V[1]-\lambda J),\\
-i\partial_s\hat\psi[1]
&=&
\hat\psi[1]\big(V[1]-\lambda J).
\ee

\medskip
{\it Remarks\/}: 
(a) The above transformation is called ``binary" because it can
be shown to be a composition of two mutually conjugated 
``elementary" Darboux-type
transformations introduced and analyzed in \cite{LU,ZL,L}.
In this respect it is similar to the standard binary Darboux
transformation discussed in \cite{MS}. 

(b) The form of the transformation resembles
the expressions occuring in the so-called dressing method
introduced by Zakharov in the context of the Riemann-Hilbert
factorization problem \cite{Z,Z1,Z2,Z3,Z4}. The difference between the binary
transformation and the dressing method lies essentially in the
form of $P$ and restrictions imposed on the parameters $\mu$ and
$\nu$. In the original Zakharov construction the parameters were
related to poles of complex functions and the whole approach 
relied on analyticity properties of objects under consideration. 
This complex-analytic framework made it difficult to work with
infinite-dimensional systems. The binary Darboux technique is
purely algebraic and therefore the above restrictions no longer
apply. The parameters are arbitrary and an extension to
infinite-dimensional systems is natural, cf. the example analyzed
in \cite{SLMC} where $J=H$ was a Hamiltonian of a
quantum-mechanical harmonic oscillator. 
 
\section{Nonlinear density matrix equations}

In this section we discuss a class of nonlinear equations that
can be solved by the binary Darboux technique described in the
previous section. Their common feature is the existence of a
Darboux-covariant Lax pair. The class is parametrized by a
natural number $n$, and for $n=1$ reduces to 
EAvNE analyzed in
\cite{SLMC}. The EAvNE for projectors reduces to the pure-state
vNE which, on the other hand, is
equivalent to the Schr\"odinger equation (SE). For $n>1$ the
equations we discus in this section are equivalent, for pure states, to
a class of nonlinear Schr\"odinger equations (NSE). 

Let $A$ be a time-independent self-adjoint operator. 
Consider the following class of nonlinear vNE's
\be
i\dot\rho
&=&
[H(\rho),\rho]\\
&=&
{\sum_{k=0}^n} [A^{n-k}\rho A^k,\rho]
={\sum_{k=0}^n} [A^{n-k},\rho A^k \rho] \label{n-eq}
\ee
where the dot denotes the time derivative. For $n=1$ and $A=H$
one finds 
\be
H(\rho)=H \rho+ \rho H
\ee
and the nonlinear vNE becomes the EAvNE
\be
i\dot\rho
=[H \rho+ \rho H,\rho]=[H,\rho^2].
\ee
Let us stress that the problem we consider is generically
infinite-dimensional and $\rho$ is positive, bounded,
trace-class and Hermitian. These constraints are
Darboux-covariant, as will become clear later, if the binary
$\nu=\bar\mu$ transformation is considered. 

The Lax representation of (\ref{n-eq}) is given by the pair of
linear equations 
\be
z_\mu |\varphi\rangle 
&=& 
(\rho-\mu A)|\varphi\rangle,\label{1a}\\
i|\dot\varphi\rangle 
&=&
\Big(
\sum_{k=0}^n A^{n-k}\rho A^k-\mu A^{n+1}\Big)|\varphi\rangle,
\label{1b}
\ee
where $z_\mu$, $\mu\in{\bf C}$. In order to use the binary
technique we have to consider the additional {\it two\/}
conjugated Lax pairs with parameters $\lambda$, $z_\lambda$,
$\nu$, $z_\nu$:
\be
z_\lambda \langle\psi| 
&=& 
\langle\psi|(\rho-\lambda A),\label{2a}\\
-i\langle\dot\psi| 
&=&
\langle\psi|\Big(
\sum_{k=0}^n A^{n-k}\rho A^k-\lambda A^{n+1}\Big),
\label{2b}\\
z_\nu \langle\chi| 
&=& 
\langle\chi|(\rho-\nu A),\label{3a}\\
-i\langle\dot\chi| 
&=&
\langle\chi|\Big(
\sum_{k=0}^n A^{n-k}\rho A^k-\nu A^{n+1}\Big).\label{3b}
\ee
In what follows the $\mu$- and $\nu$-pairs (\ref{1a})--(\ref{1b}) 
and (\ref{3a})--(\ref{3b}) will be used to
define the binary transformation of the conjugated
$\lambda$-pair (\ref{2a})--(\ref{2b}).

\medskip
\noindent
{\bf Lemma~1.} The Lax pair (\ref{2a})--(\ref{2b}) is
covariant under the binary Darboux transformation.

\medskip
\noindent
{\it Proof\/}: The Lax pair can be regarded as a particular case
of the zero-curvature representation 
\be
-i\langle\psi'| 
&=& 
\langle\psi|(\rho-\lambda A)
=:
\langle\psi|(V_1-\lambda J_1)
,\label{4a}\\
-i\langle\dot\psi| 
&=&
\langle\psi|\Big(
\sum_{k=0}^n A^{n-k}\rho A^k-\lambda A^{n+1}\Big)\nonumber\\
&=:&
\langle\psi|(V_2-\lambda J_2),
\label{4b}
\ee
where the prime denotes a derivative with respect to some
auxiliary parameter $\tau$, and $\rho'=0$. Eqs.~(\ref{4a}), (\ref{4b}) are
examples of general Zakharov-Shabat problems. The binary Darboux
transformation implies 
\be
-i\langle\psi[1]'| 
&=& 
\langle\psi[1]|\big(V_1[1]-\lambda J_1\big),\\
-i\langle\dot\psi[1]| 
&=&
\langle\psi[1]|\big(
V_2[1]-\lambda J_2\big),
\ee
where
\be
V_1[1] &=& V_1+(\mu-\nu)[P,J_1]\\
 &=&
\rho+(\mu-\nu)[P,A]=:\rho[1]\\
V_2[1] &=& V_2+(\mu-\nu)[P,J_2]\\
&=&
\sum_{k=0}^n A^{n-k}\rho A^k
+(\mu-\nu)[P,A^{n+1}]\nonumber\\
&=&
\sum_{k=0}^n A^{n-k}\rho A^k
+(\mu-\nu)\sum_{k=0}^n A^{n-k}[P,A]A^k\nonumber\\
&=&
\sum_{k=0}^n A^{n-k}\rho[1] A^k.
\ee
The $\tau$-dependence of the solutions of the zero-curvature
equations is given by
\be
|\varphi(t,\tau)\rangle &=&e^{-iz_\mu \tau}
|\varphi(t,\tau=0)\rangle\\
\langle\chi(t,\tau)| &=&e^{iz_\nu \tau}
\langle\chi(t,\tau=0)|\\
\langle\psi(t,\tau)| &=&e^{iz_\lambda \tau}
\langle\psi(t,\tau=0)|
\ee
so that $P$ is $\tau$-independent and, hence, 
\be
\langle\psi[1](t,\tau)| &=&e^{iz_\lambda \tau}
\langle\psi[1](t,\tau=0)|.
\ee
Therefore
\be
-i\langle\psi[1]'| 
&=& 
\langle\psi[1]|(\rho[1]-\lambda A)=z_\lambda \langle\psi[1]|\\
-i\langle\dot\psi[1]| 
&=&
\langle\psi[1]|\Big(
\sum_{k=0}^n A^{n-k}\rho[1] A^k-\lambda A^{n+1}\Big)
\ee
which was to be proved. $\Box$

\medskip
{\it Remarks\/}: (a) The above Lemma implies that
Eqs.~(\ref{n-eq}) are invariant under the action of the binary
Darboux transformation, since the stationarity conditions imply
$P'=0$ and $\rho[1]'=0$ if $\rho'=0$. 

(b) For $\nu=\bar\mu$ $\rho=\rho^{\dag}$
implies $\rho[1]=\rho[1]^{\dag}$. 

(c) $\Tr\rho=\Tr\rho[1]$. 

(d) For pure states, i.e. $\rho=|\Psi\rangle\langle\Psi|$,
$\langle\Psi|\Psi\rangle=1$, Eqs.~(\ref{n-eq}) are equivalent to a
class of nonlinear SE. Indeed, in this case 
$$
\rho A^k
\rho=|\Psi\rangle\langle\Psi|A^k|\Psi\rangle\langle\Psi|= 
\Tr(\rho A^k)\rho
$$
and
\be
i\dot\rho
&=&
{\sum_{k=0}^n} [A^{n-k},\rho A^k \rho]
=
{\sum_{k=0}^n}\Tr(\rho A^k) [A^{n-k},\rho].\label{pure}
\ee
The solutions of (\ref{pure}) are 
\be
\rho(t)=e^{-i {\sum_{k=0}^n}\Tr(\rho A^k) A^{n-k} t}
\rho(0)
e^{i {\sum_{k=0}^n}\Tr(\rho A^k) A^{n-k} t}\nonumber
\ee
since $\Tr(\rho A^k)$ are time-independent. 
The same dynamics could be obtained from the NSE 
\be
i|\dot \Psi\rangle=\sum_{k=0}^{n-1}\langle\Psi|A^k|\Psi\rangle
 A^{n-k}|\Psi\rangle.\label{nlSE}
\ee
Let us recall that the case $n=1$, $A=H$ is equivalent to the
EAvNE which for pure states gives the ordinary vNE. This property
is seen also in (\ref{nlSE}) since then the corresponding SE is
the ordinary linear one.

\section{Binary Darboux transformation as a similarity
transformation} 

Spectra of all the solutions of the EAvNE discussed in
\cite{SLMC} were {\it invariant\/} under the binary
transformation. 
Below we show that the result is not a pecularity of the
EAvNE but holds for all operator equations that are 
compatibility conditions for a Lax pair of ZS equations. In a
more general situation where instead of a Lax pair we have a
zero-curvature pair, the result does not have to hold. 
For solutions that are not self-adjoint the result extends
separately to both the left and the right spectra. 

Consider the following three general zero-curvature pairs 
\be
i |\varphi'\rangle 
&=& 
(\rho-\mu A)|\varphi\rangle=z_\mu |\varphi\rangle ,\label{L1a}\\
i|\dot \varphi\rangle 
&=&
\big(V(\rho) - \mu J\big)|\varphi\rangle,\label{L1b}\\
-i\langle\chi'| 
&=& 
\langle\chi|(\rho-\nu A)=z_\nu \langle\chi| \label{L3a}\\
-i\langle\dot \chi|
&=&
\langle\chi| \big(V(\rho) - \nu J\big),\label{L3b}\\
-i\langle\psi'| 
&=& 
\langle\psi|(\rho-\lambda A)\label{L2a}\\
-i\langle\dot \psi|
&=&
\langle\psi| \big(V(\rho) - \lambda J\big),\label{L2b}
\ee
The only assumption we make about $\rho$, $J$, $A$, and 
$V(\rho)$ is the covariance of
(\ref{L2a})--(\ref{L2b}) under the binary Darboux transformation
constructed with the help of $\langle\chi|$ and $|\varphi\rangle $. 
Let us note that the stationarity with respect to $\tau$ is
assumed only for the first two pairs. 

\medskip
\noindent
{\bf Theorem~2.} Under the above assumptions the binary Darboux
transformation $\rho\mapsto\rho[1]$ is a similarity
transformation,
$\rho[1]=T\rho T^{-1}$, where
\be
T &=& {\bbox 1}+\frac{\mu-\nu}{\nu}P=e^{P\ln\frac{\mu}{\nu}}
\ee
{\it Proof\/}: 
By definition $\rho[1]=\rho+(\mu-\nu)[P,A]$. 
Eqs.~(\ref{L1a}), (\ref{L3a}) imply 
\be
z_\mu P
&=& 
(\rho-\mu A)P\label{LP1a}\\
z_\nu P
&=& 
P(\rho-\nu A)\label{LP3a}
\ee
from which it follows that 
\be
P(\rho-\mu A)P
&=& 
(\rho-\mu A)P\label{PLP1a}\\
P(\rho-\nu A)P
&=& 
P(\rho-\nu A).\label{PLP3a}
\ee
Multiplying (\ref{PLP1a}) by $\nu$, (\ref{PLP3a}) by $\mu$, and
subtracting the resulting equations we get 
\be
[P,A]=\frac{\nu-\mu}{\mu\nu} P\rho P-\frac{1}{\mu} \rho P+\frac{1}{\nu}
P\rho. 
\ee
Inserting this expression into the definition of $\rho[1]$ we
obtain 
\be
\rho[1]=
\Big({\bbox 1}+\frac{\mu-\nu}{\nu}P\Big)
\rho
\Big({\bbox 1}+\frac{\nu-\mu}{\mu}P\Big).\label{central}
\ee
To see that the operators occuring at both sides of $\rho$ are
inverses of each other it is sufficient to use
$P^2=P$. For any $z\in\bbox C$ and any $P$ satisfying $P^2=P$
one finds 
$$
\exp(zP)=\bbox 1-P+e^zP
$$
and, in particular, 
$$
e^{P\ln\frac{\mu}{\nu}}=\bbox 1-P+\frac{\mu}{\nu}P=T.
$$
$\Box$ 

\medskip
{\it Remarks\/}: (a) For Hermitian $P$ and $\nu=\bar\mu$ the
operator $T$ is unitary. If in addition $\mu$ is imaginary then 
$T=\bbox 1 -2P$. 

(b) The assumptions we have made can be weakened since the proof
is based on the equalities (\ref{PLP1a}) and
(\ref{PLP3a}) which are weaker than (\ref{L1a}), (\ref{L3a}).

(c) The proof extends without any modification to the operator
version of the binary Darboux transformation, where we take
the operators $\hat\varphi$ and $\hat\chi$ instead of the
Hilbert-space vectors $|\varphi\rangle$ and $\langle\chi|$, and 
$P=\hat\varphi(p\hat\chi\hat\varphi p)^{-1}\hat\chi$. 

(d) The result applies to all the equations described in the
previous section. As a consequence all their solutions $\rho[1]$
obtained from a density matrix $\rho$ with $\nu=\bar\mu$ are
also density matrices. 

\section{Spectrum shifting and rescaling}

It is often convenient to
start with a solution that is not a density matrix. For example,
one can start with a $\rho$ which is not positive, or whose trace is
not equal to 1. To turn it into a density matrix one first has to add
an appropriate positive operator to $\rho$ and then renormalize the
solution to get $\Tr\rho=1$. Below we give examples of such
transformations. 

Take an operator $X$ satisfying $[X,A]=[X,\rho]=0$. Such an $X$
always exists (e.g. $X=\Lambda \bbox 1$, $\Lambda\in \bbox R$). 
Assume $\rho(t)$ is a solution of (\ref{n-eq}). Then 
\be
\rho_X(t)
&=&
e^{-i(n+1)XA^nt}\big(\rho(t)+X\big)e^{i(n+1)XA^nt}
\ee
also satisfies (\ref{n-eq}). 
In particular, for $X=\Lambda \bbox 1$ the spectrum of $\rho_X$
is shifted by $\Lambda$ with respect to this of $\rho$. 
For $n=1$ one obtains the gauge transformation of the EAvNE
discussed in \cite{SLMC}. This trick is
very useful in practical computations of finite dimensional cases and
typically leads to positive but non-normalized solutions. 

To change the normalization one uses the rescaling symmetry of
(\ref{n-eq}): 
\be
\rho(t)\mapsto \rho_Y(t)=Y\rho(Yt), \quad Y\in \bbox R.
\ee
As an application of the above two transformations consider the
problem of stationary seed solutions $\rho$. The solutions discussed
in \cite{SLMC} were starting with non-stationary $\rho$'s, since those
satisfying $[\rho,A]=0$
lead to the projector $P$ commuting with both $A$ and $\rho$, and the 
binary transformation is trivial. Still, there exists another 
class of stationary solutions of (\ref{n-eq}), obtained if $A\rho=-\rho A$. 
Now the projector $P$ will not, in general, commute with $\rho$ and
$A$, and the binary transformation may be nontrivial. 

Assume $[P,\rho]\neq 0$. 
For $n=2m$ one has $\sum_{k=0}^n (-1)^k=1$ and
\be
i|\dot\varphi\rangle 
&=&
\Big(
A^n\rho-\mu A^{n+1}\Big)|\varphi\rangle=
z_\mu A^n|\varphi\rangle\label{15}
\ee
whose formal solution is 
\be
|\varphi(t)\rangle = e^{-iz_\mu A^n t}|\varphi(0)\rangle \label{sol even}.
\ee
For odd $n$ we get $\sum_{k=0}^n (-1)^k=0$ and 
\be
i|\dot\varphi\rangle 
&=&
-\mu A^{n+1}|\varphi\rangle
\ee
implying 
\be
|\varphi(t)\rangle = e^{i\mu A^{n+1} t}|\varphi(0)\rangle\label{sol odd} .
\ee
The fact that $\mu$ and $z_\mu$ are complex results, in general,
in a nontrivial time evolution of $P$ arising from the nontrivial
contribution from the time
dependence of $\langle\varphi(t)|\varphi(t)\rangle$. 
Let us note here that the ``phase transitions" found for the
Euler-Arnold-von Neumann equation had a similar origin although the
``seed solutions" $\rho$ satisfied a different condition. An exceptional
situation occurs if the solution of the Lax pair turns out to be an eigenstate 
of $A^{2m}$ (for $n=2m$) or $A^{2(m+1)}$ (for $n=2m+1$). 
Let us also note that $\rho$ that anticommutes with $A$ does commute with 
$A^2$ and therefore also with the generators of the time evolution 
given by (\ref{sol even}) and (\ref{sol odd}). This is an important property
allowing to look for an eigenstate of $\rho - \mu A$ at $t=0$. 

For $\rho$ anticommuting with $A$ one finds $\Tr
\rho A=0$. This implies that $\rho$ and $A$ cannot be simultaneously 
positive. In a finite
dimensional case one can easily shift the spectrum of $\rho$ by a
number and then renormalize the resulting solution by means of the
above two transformations, and in this way use $\rho$ as an
intermediate step towards a density matrix $\rho[1]$ with positive $A$. 
An analysis of explicit 
solutions obtained by this technique will be presented in a forthcoming
paper \cite{MCMSKW}.

\section{Further perspectives}

The central result of this Letter is the formula (\ref{central}). For
$\mu\neq\bar\nu$ the operator $T$ is nonunitary. A possibility that
has not been explored as yet is to look for Hermitian $\rho[1]$ but
starting with a non-Hermitian $\rho$ and nonunitary $T$. In still
more general perspective, it is clear that the method we have
developed works for a class of nonlinear operator equations 
of the Heisenberg form $i\dot F=[H(F),F]$ and, hence, may find
applications in quantum optics and field theory. 
The technique is well suited for solving infinite-dimensional
problems, but a practical complication is associated with constraints
typical of density matrices. Hermiticity is an easy issue here, but
real difficulties occur if one looks for trace-class and positive
solutions. 

Another open question is how to solve vNE's of the general form 
\be
i\dot\rho=[H,f(\rho)].
\ee
There are reasons to investigate this type of nonlinearities 
since there exist links of $f(\rho)=\rho^q$, $q>0$, 
to nonextensive $q$-statistics (cf. \cite{MCJN} and references
therein). At the moment no binary-Darboux covariant Lax pairs are
known for $q\neq 2$. 
Actually, looking more closely at the structure of Zakharov-Shabat
equations and properties of the binary transformation 
one can conjecture that this particular Darboux transformation can
work only for vNE's with Hamiltonians {\it linear\/} in $\rho$. 
For example, we are aware of a Lax pair leading to 
\be
i\dot\rho=[H,\rho^3]=[\rho^2 H+\rho H \rho +H \rho^2,\rho]
=
[H(\rho),\rho]
\ee
but it is not covariant under the binary transformation: The reason
is the quadratic dependence on $\rho$ of the Hamiltonian 
$H(\rho)$. On the other
hand, all $f(\rho)$ nonlinearities with polynomial $f$ have similar
properties when looked at from a Lie-Poisson perspective (say, the same
Casimirs and other invariants) and there is basically no reason why
these equations should not be solvable by a Darboux-type technique.  
It is possible that such transformations should be looked for between
those obtained by a composition of {\it three\/} or more elementary
transformations discussed in \cite{ZL,L}.

A new class of constraints, which have not been
considered so far, are those related to commutation relations. To give
an example, in quantum optics the task is often reduced to solving 
nonlinear Heisenberg equations for creation and annihilation
operators. Assuming the problem is treated by a Darboux technique,
the constraint will be of the form $[a[1],a[1]^{\dag}]=1$ and the
fact that we are dealing with a similarity transformation
(\ref{central}) may be of crucial importance. 

All these problems are a subject of current study. 

\begin{acknowledgments}
Our work was supported by the KBN Grant No. 2~P03B~163~15. The
work of M.K. and M.C. was financed in part by the Polish-Flemish 
Grant No. 007. This work was done partly during our stays in
Brussels, Antwerp, and Clausthal. We thank J.~Naudts for fruitful
discussions and the organizers of the Antwerp workshop for
financial support. M.C. also thanks Alexander von Humboldt-Stiftung
for making possible his stay in ASI where the final version of the paper was
completed.  
\end{acknowledgments}

\end{document}